\documentclass[aps,prl,twocolumn,showpacs,floatfix,superscriptaddress]{revtex4-1}

\pdfoutput=1

\usepackage{amsmath}
\usepackage{amssymb}
\usepackage{graphicx}
\usepackage{epstopdf}
\usepackage{latexsym}

\begin{document}

\title{Orbital-FFLO state in a chain of high spin ultracold atoms}
\author{E. Szirmai}
\affiliation{BME-MTA Exotic Quantum Phases Research Group, Institute of Physics, Budapest University of Technology and Economics, 
Budafoki \'ut 8., H-1111 Budapest, Hungary}

\begin{abstract}
Recent experiments with Yb-173 and Sr-87 isotopes provide new possibilities to study high spin two-orbital systems. Within these experiments part of the atoms are excited to a higher energy metastable electronic state mimicking an additional internal (orbital) degree of freedom. The interaction between the atoms depends on the orbital states, therefore four different scattering channels can be identified in the system characterized by four independent couplings. When the system is confined into a one-dimensional chain the scattering lengths can be tuned by changing the transverse confinement, and driven through four resonances. Using the new available experimental data of the scattering lengths we analyze the phase diagram of the one-dimensional system as the couplings are tuned via transverse confinement, and the populations of the two orbital states are changed. We found that three orders compete showing power law decay: a state with dominant density wave fluctuations, another one with spin density fluctuations, and a third one characterized by exotic Fulde-Ferrell-Larkin-Ovchinnikov-like pairs consisting one atom in the electronic ground state and one in the excited state. We also show that sufficiently close to the resonances the compressibility of the system starts to diverge indicating that the emerging order is unstable and collapses  to a phase separated state with a first order phase transition.
\end{abstract}

\pacs{67.85.-d, 03.75.Ss, 72.15.Nj}
\date{\today}
\maketitle

In 2010 Gorshkov and collaborators \cite{Gorshkov2010} proposed a fundamental model to describe two-orbital physics within ultracold atomic experiments. In the proposed setup alkaline-earth atoms are loaded into an optical lattice and part of the atoms are optically driven to their metastable excited electronic state ($^3P_0$) $|e\big>$, while the rest of the atoms remains in the electronic ground state ($^1S_0$) $|g\big>$. Accordingly, the roles of the two orbital states are played by the electronic states of the atoms. The atoms interact via weak Van der Waals interaction that does not depend on the spin degree of freedom~\cite{Lewenstein2007,Bloch2008, Ketterle2008a,Lewenstein2012, Cazalilla2014}, but couples the two orbital states. This leads to an SU($2F+1$) symmetric interaction between atoms with hyperfine spin $F$, and the strength of the interaction depends only on the artificial orbital states of the scattering atoms. Such two-orbital high spin systems have been realized experimentally in 2014 by three independent groups: the preparation of an SU(6) system with Ytterbium-173 isotopes ($F=5/2$) has been reported in Refs.~\cite{Cappellini14a,Scazza14a}, and an SU(10) symmetric system with Strontium-87 isotopes ($F=9/2$) in Ref.~\cite{Zhang14a}. These experiments open a new direction to study high spin fermionic systems~\cite{Honerkamp2004,Szirmai2006,Greiter2007,Paramekanti2007,Szirmai2008,Cazalilla2009,Hermele2009,Wu2010,Taie2010,Desalvo2010,Xu2010,Manmana2011,Szirmai2011a,Szirmai2011,Krauser2012,Zhang2014,Pagano2014a} with additional degree of freedom, especially when the atoms are sufficiently strongly confined into a one-dimensional tube. Assuming harmonic confinement with transverse vibrational energy $\omega_\perp$, the interaction strength ($g$) can be tuned~\cite{Olshanii1998,Petrov2001,Peano2005,Bloch2008,Pricoupenko2008,Haller2010,Dunjko2011,Zhang2011} by changing the transverse confinement as
\begin{equation}
\label{eq:coupling}
g=\frac{2 {\hbar}^2}{m} \frac{1}{a_\perp^2}\frac{a_{3D}}{1 - b \frac{a_{3D}}{a_\perp}},
\end{equation}
where $a_{3D}$ is the scattering length measured in a three dimensional system, $b \approx 1.4603$ is a numerical constant, and $a_\perp = \sqrt{\hbar/m\omega_\perp}$ is the transverse oscillator length in case of atomic mass $m$. From Eq.~\eqref{eq:coupling} it is readily seen that confinement can induce a resonance of the one-dimensional scattering length at $a_\perp\approx b \cdot a_{3D}$ opening a possibility to change the interaction even from the repulsive side ($g>0$) of the interaction to the attractive one ($g<0$).

In an earlier work \cite{Szirmai13a} we predicted the possible existence of an exotic pairing, an orbital-FFLO (Fulde, and Ferrell~\cite{Fulde64a} and Larkin, and Ovchinnikov~\cite{Larkin64a}) state for $^{87}$Sr isotope based on the current available data for the scattering lengths --- which estimated data have turned out to be completely wrong, even the sign is mismatching in some cases. Here we show that the orbital-FFLO state is very robust, and a chain of two-orbital $^{87}$Sr isotopes just as $^{173}$Yb atoms can in fact be driven to a state of the exotic orbital-FFLO. Accordingly, this exotic pairing state can be realized within an ultracold atomic experiment. Note that the SU(N) symmetry is not violated in these systems, therefore the expectation values of the orbital-FFLO order parameters are necessarily zero --- in agreement with the statement made in Ref.~\cite{Bois2016} for similar system. Nevertheless, we found that the correlation functions shows algebraic decay leading to an unusual spin liquid-like pairing state~\footnote{Similar spin liquid states can be observed e.g. in the $J_1$-$J_2$ isotropic Heisenberg chain which is characterized by quasi-long range antiferromagnetic correlations with zero local spin expectation values~\cite{Fazekas1999}.}. We also show that 
for sufficiently strong interactions the system collapses into a finite region of the chain indicating by the divergence of the sound velocity of the density oscillations. Such segregation (see eg.~\cite{voit92a}) can occur only if the effective masses of the two orbits are imbalanced, or if the $g$ couplings which characterize the interaction strengths between the atoms in the two orbital states are not equal. The effective mass imbalance can be tuned e.g. via the population of the two orbits. On the other hand, the latter requirement (the orbital state dependence of the $g$ couplings) is always fulfilled, since the four scattering lengths are different. Therefore, close to the resonances the atom cloud always becomes segregated.
Using the now available experimental data of all the four scattering lengths of the two-orbital systems in case of $^{173}$Yb and also $^{87}$Sr isotopes, respectively, we present --- within hydrodynamical treatment~\cite{Vondelft1998,Gogolin2004,Giamarchi2004} using semiclassical approach --- the complete phase diagram of such two-orbital systems when they are confined into a one-dimensional chain. We show that the number of the spin components ($2F+1$) does not play significant role in the low energy phase diagram, the existence of the orbital-FFLO state is not sensitive to the specific value of $F$, and the two isotopes exhibit very similar characteristic features.

In order to fit the experimental data into the model parameters let us have a look on the effective Hamiltonian. Due to the optical pumping the two orbital states are off-resonant, and the orbital dependent kinetic terms are decoupled~\cite{Gorshkov2010}: 
$H_{0}^{\alpha}=-\sum_{i,\sigma} t_\alpha ( c_{i,\alpha,\sigma}^\dagger c_{i+1,\alpha,\sigma} + H.c.)$, with $\alpha=g$ (electronic ground state) or $e$ (electronic excited state). Here and in the following $c_{i,\alpha,\sigma}^\dagger$ ($c_{i,\alpha,\sigma}$) creates (annihilates) an atom in the orbital state $\alpha$ with spin $\sigma$ on site $i$. Assuming that the low energy excitations determine the main properties of the system, the tight-binding spectrum can be linearized around the Fermi points leading to the Fermi velocities: $\hbar v_{g(e)} = 2 t_{g(e)} a\, \sin(k_{g(e)} a )$, where $a$ is the size of the unit cell along the one-dimensional optical lattice, and $k_{g(e)}$ is the Fermi momentum determined by the population $N_{g(e)}$ of the corresponding electronic states: $k_{g(e)} = \pi f_{g(e)} = \pi N_{g(e)}/(2F+1)L$, and $L$ is the length of the chain.  The interaction between the atoms can be described by density-density and orbital-exchange scatterings:
$H_\mathrm{int} =  \frac12 \sum_{i, \sigma , \sigma'} \Big[ \sum_{\alpha=g,e} g_\alpha n_{i,\alpha,\sigma}n_{i,\alpha,\sigma'} +
 g_{ge}\, n_{i,e,\sigma}n_{i,g,\sigma'} + g_{ge}^{\textrm{ex}} \,
 c_{i,g,\sigma}^\dagger c_{i,e,\sigma'}^\dagger c_{i,g,\sigma'} c_{i,e,\sigma} \Big]$, where $n_{i,\alpha,\sigma}=c_{i,\alpha,\sigma}^\dagger c_{i,\alpha,\sigma}$ is the particle number operator.
The interorbital scattering length has different values depending on whether the two-particle state of the two colliding atoms is symmetric ($g_{ge}^+$) or antisymmetric ($g_{ge}^-$) for the exchange of the two orbital states. Their linear combinations drive the density-density interaction: $g_{ge}=g_{ge}^+ + g_{ge}^-$, and the orbital exchange interaction: $g_{ge}^{\textrm{ex}}=g_{ge}^+ - g_{ge}^-$. Table~\ref{tab:measured_as}. shows the available experimental data of the various scattering lengths for $^{87}$Strontium~\cite{Zhang14a} and for $^{173}$Ytterbium~\cite{Cappellini14a,Scazza14a}. Since there is no Feshbach resonance available for these systems, in order to drive the system into resonance the confinement energy $\omega_\perp$ has to be tuned.  Typically the transverse confinement energy $\omega_\perp/2 \pi$ is in the order of 100~kHz~\cite{Bloch2008} realizing a one-dimensional system. However, its specific value depends on the trapped atom and the details of the experimental setup. In case of $^{173}$Yb the transverse confinement can be tuned in the range of $700-1400$ $a_0$ \footnote{G. Pagano and L. Fallani, private communication.}, further tuning needs specific efforts. From the above value of $\omega_\perp$ the transverse oscillation length is $a_\perp \approx 680\,\, a_0$ in case of $^{87}$Sr.
Therefore, in order to reach most resonances the transverse oscillation length has to be decreased, i.e. stronger confinement is needed. Such decreasing can in principle be done by changing the lattice wavelength or the lattice depth (see e.g.~\cite{Bloch2008}). 

\begin{table}
\begin{tabular}{ | c | c | c || c | c |  }
\hline
&\multicolumn{2}{ | c || }{ $^{87}$Sr } & \multicolumn{2}{ | c | }{ $^{173}$Yb }  \\ 
\cline{2-5}
[$a_0$] & $a_{3D}$  & $b \cdot a_{3D}$ & $a_{3D}$ & $b \cdot a_{3D}$ \\
\hline
 \hspace{0.3cm} $a_g$  \hspace{0.3cm} &   \hspace{0.3cm} 96.2 \hspace{0.3cm} &  \hspace{0.3cm} 140.5 \hspace{0.3cm} & \hspace{0.3cm} 199.4  \hspace{0.3cm} &  \hspace{0.3cm} 291.2 \hspace{0.3cm}  \\
 $a_e$ & 176  & 257 & 306.7  & 447.9 \\
 $a_{ge}^+$  & 169  & 246.8 & 219.5 & 320.5 \\
 $a_{ge}^-$   & 68  & 99.3 & 3300  & 4819 \\
\hline
\end{tabular}
\caption{The scattering length for $^{87}$Sr and $^{173}$Yb, respectively \cite{Cappellini14a,Scazza14a,Zhang14a}. Resonance is expected when the transverse oscillator length $a_\perp$ is around $b \cdot a_{3D}$. }
\label{tab:measured_as}
\end{table}

\begin{figure}
\includegraphics[scale=0.58]{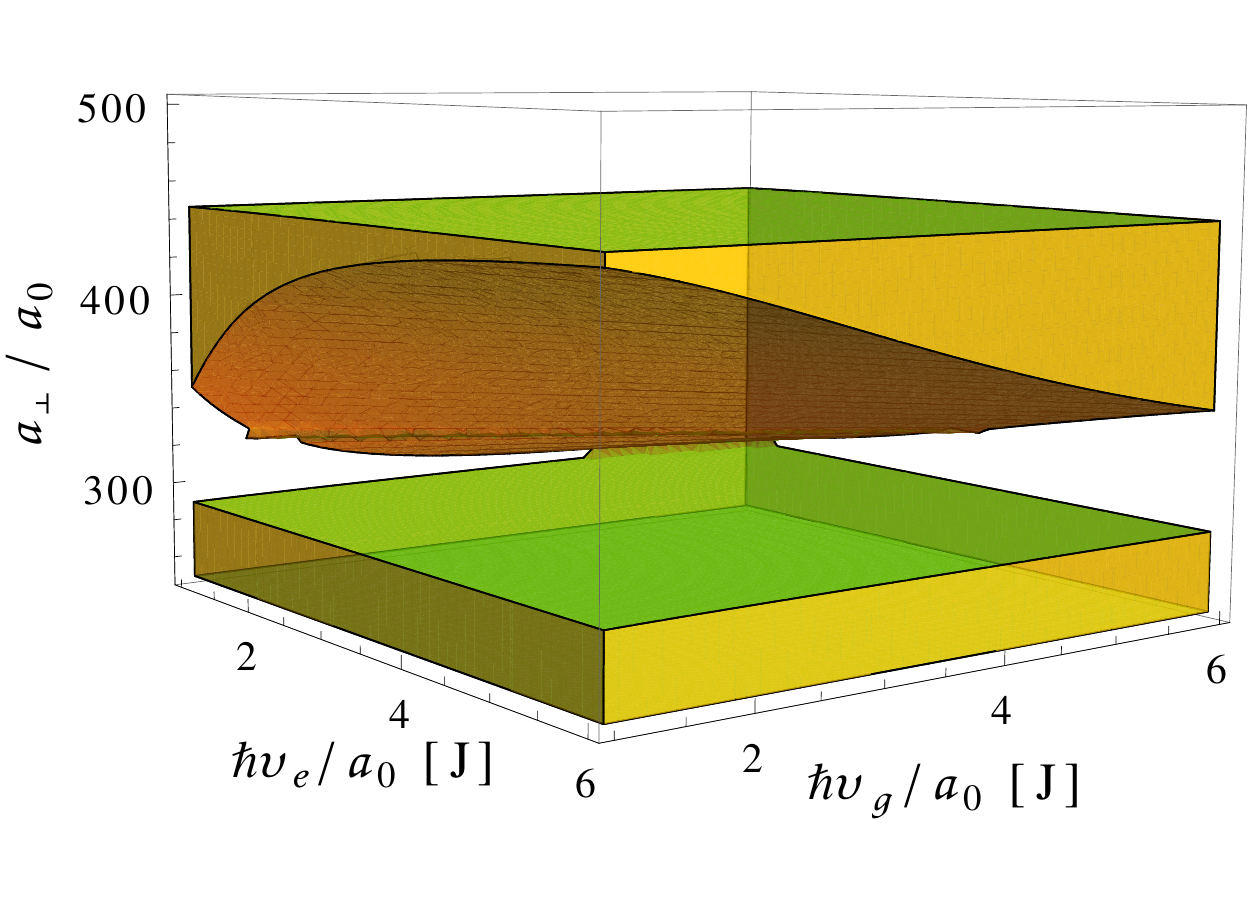}
\includegraphics[scale=0.58]{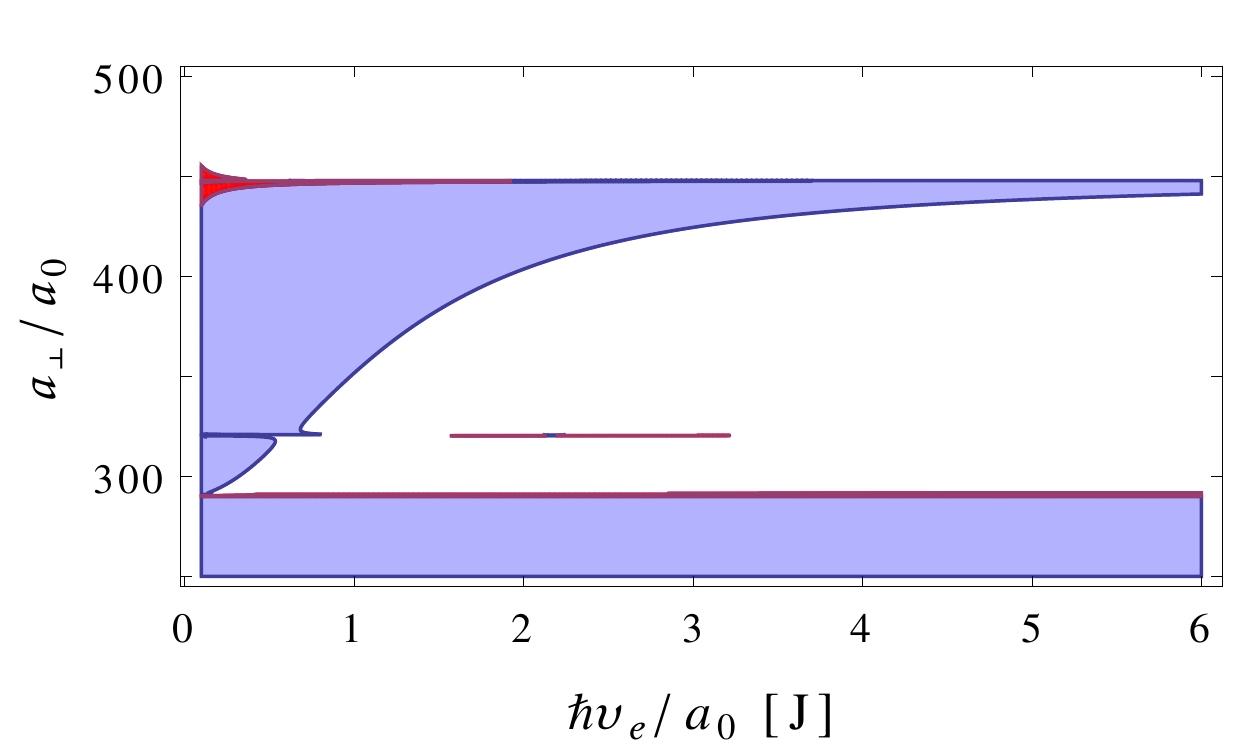}
\caption{(Color online) Upper panel: Phase diagram of a two-orbital Yb-173 cloud as a function of the tunneling in the electronic ground state $v_g$ and excited state $v_e$, and the transverse confinement $a_\perp$. The filled volume indicates the parameter region where the orbital-FFLO suppresses the density wave fluctuations, while in the white region the density fluctuations dominate. Lower panel:
Intersection of the phase diagram in the upper panel at $\hbar v_g / a_0 = 1$ Hz. The blue (gray) area denotes the dominant orbital FFLO state with subdominant density wave, and vice versa in the white region. In the red (dark gray) region the dominant fluctuation is still the density wave but with subdominant spin density wave. }
\label{fig:phase_diagYb}
\end{figure}

\begin{figure}
\includegraphics[scale=0.58]{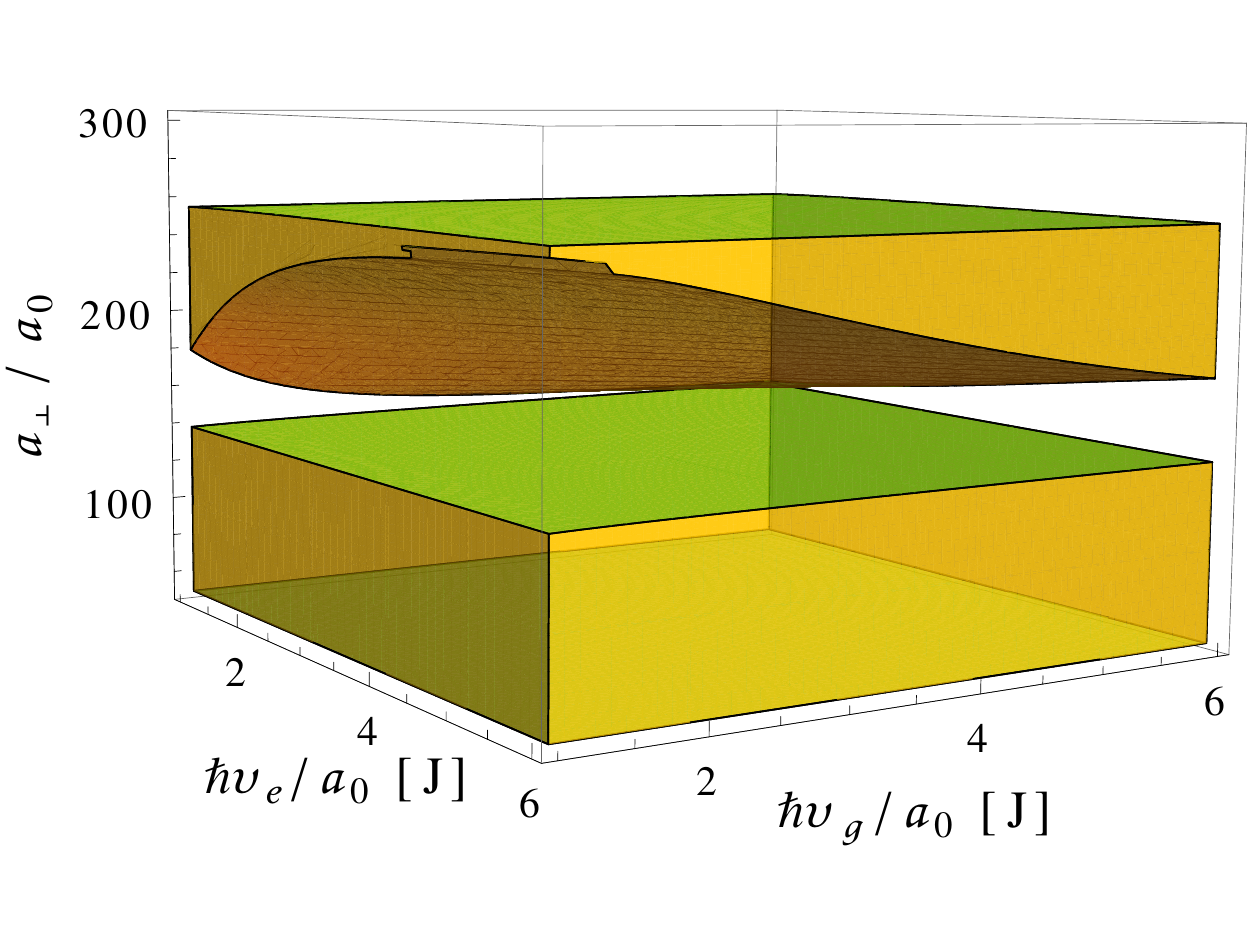}
\includegraphics[scale=0.58]{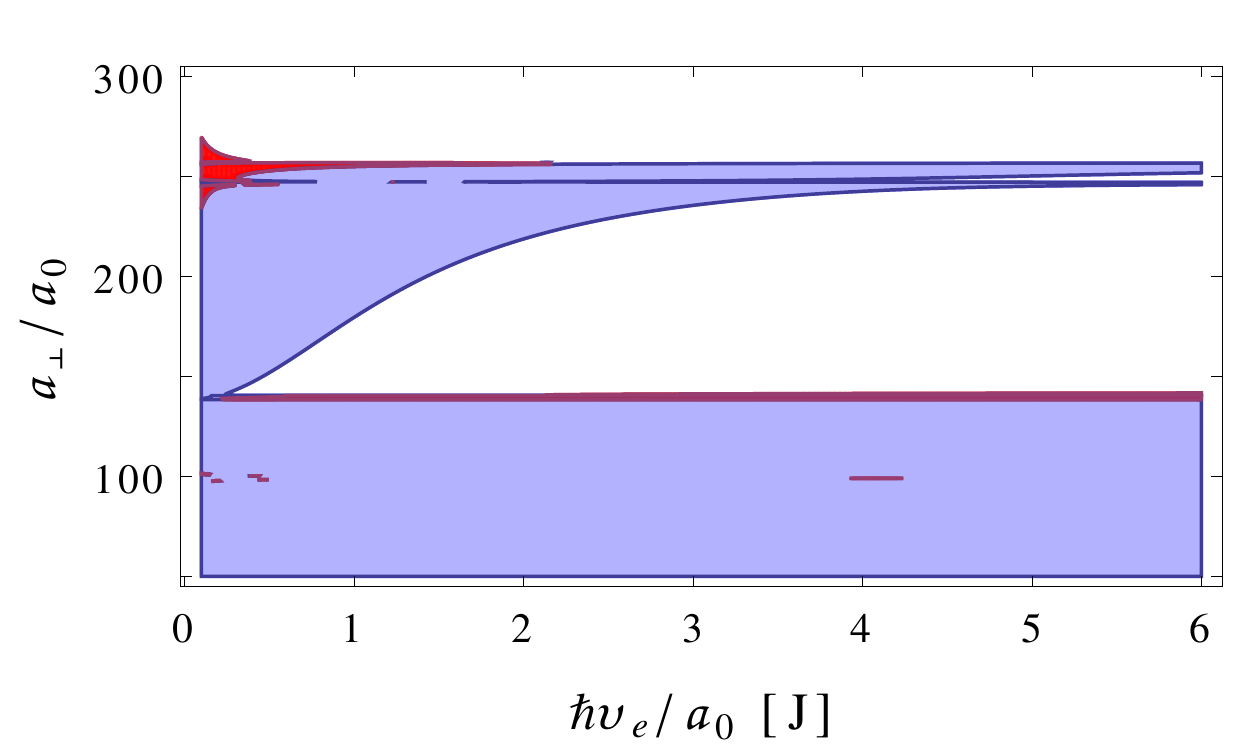}
\caption{(Color online) The same as Fig.~\ref{fig:phase_diagYb} for Strontium-87, which isotope has a hyperfine spin $F=9/2$. }
\label{fig:phase_diagSr}
\end{figure}

Exploiting our earlier hydrodynamical analysis for general SU(N) symmetric two-orbital systems~\cite{Szirmai13a} we determined the dominant fluctuations in a one-dimensional $^{87}$Sr, and $^{173}$Yb cloud, respectively, consisting atoms in the electronic ground state $^1S_0$ and in the metastable excited state $^3P_0$. In the relevant parameter regime we found three competing states which can be characterized by the quasi-long-range oscillation showing the slowest power law decay. On one hand atomic density oscillations dominate (white regions in Fig.~\ref{fig:phase_diagYb} and \ref{fig:phase_diagSr}), decaying with distance as $|r|^{-\Delta_+^\kappa - \Delta_-^\kappa}$, with exponent $\Delta_{\pm}^\kappa = \frac{1}{16\pi} ( \sqrt{\kappa_{g}} \pm \sqrt{\kappa_{e}})^2 / (1 \pm g_c^{ge} \sqrt{\kappa_{g}\kappa_{e}} ) $. Here $\kappa_{g(e)}=K_{g(e)}/u_{g(e)}$ is the compressibility of the cloud when all the particles are in the electronic ground (excited) state, $K_{g(e)}$ is the corresponding Luttinger parameter of the low energy effective theory, and $u_{g(e)}$ is the sound velocity. Their relation to the original model parameters are $K_{g(e)}= \sqrt{\frac{2\pi\hbar v_{g(e)} }{2\pi\hbar v_{g(e)} + 4F g_{g(e)} } }$, $u_{g(e)}= \sqrt{(2\pi\hbar v_{g(e)} + 2F g_{g(e)})^2 - 4F^2 g_{g(e)}^2}$, and  $g_c^{ge} =  Fg_{ge}^+ + (F+1) g_{ge}^- $ with the hyperfine spin $F=9/2$ for $^{87}$Sr, and 5/2 for $^{173}$Yb. 

\begin{figure}
\includegraphics[scale=0.58]{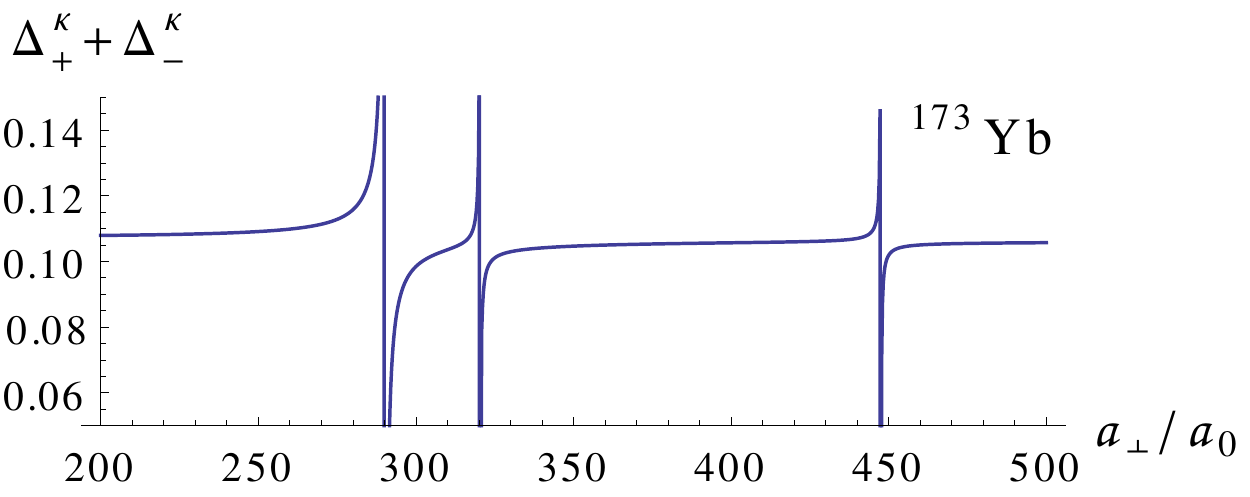}
\vskip 0.5cm
\includegraphics[scale=0.58]{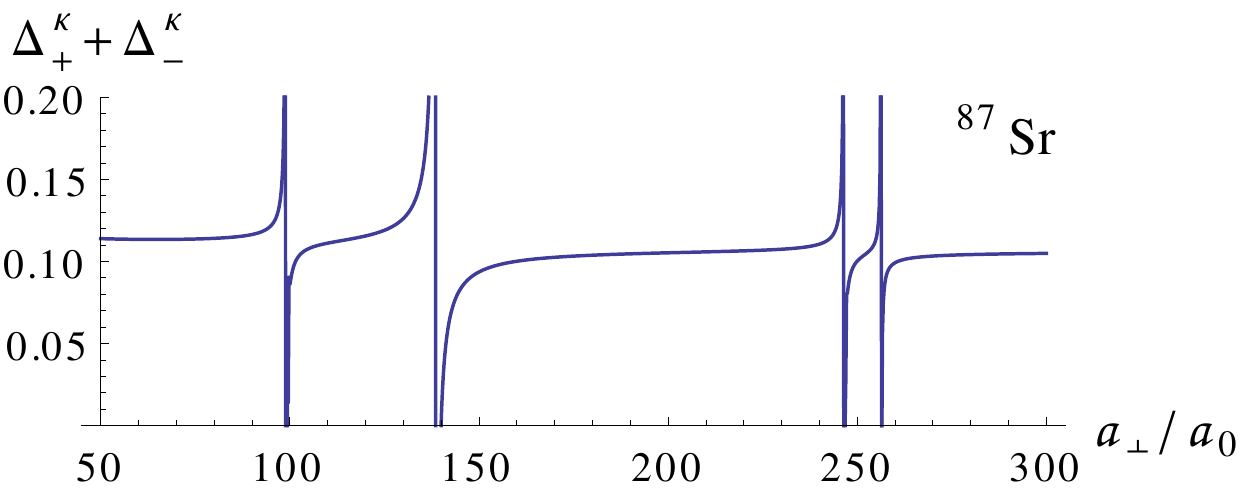}
\caption{(Color online) Critical exponent $\Delta_+^\kappa + \Delta_-^\kappa$ of the density wave order parameter as a function of the transverse confinement $a_\perp$. The tunneling in the excited state is fixed as $\hbar v_e/a_0=3$ J, and the ground state tunneling is $\hbar v_g/a_0=1$ J. }
\label{fig:crit_exp}
\end{figure}

Sufficiently far from all resonances an unusual pairing state competes with the atomic density oscillations. This state consists pairs of atoms: one from the electronic ground state and the other one from the excited state. These pairs have finite momentum in the order of the difference of the two Fermi momentums: $\pm(k_\mathrm{F}^g-k_\mathrm{F}^e)$ which is proportional to the population imbalance in the two electronic states. Accordingly, the emerging pairing state is an orbital analogue of the celebrated FFLO state. The two atoms can not form an SU(6) or SU(10) singlet, they carry spin, therefore in the SU(N) symmetric ground state they have zero expectation value. However, their correlations are invariant under SU(N) rotations, and the state preserves the global SU(N) symmetry despite their algebraic decaying. The correlations of the orbital-FFLO order parameter decay as $|r|^{-\Delta_+^\sigma - \Delta_-^\sigma}$, where the exponent is $\Delta_{\pm}^\sigma = \frac{1}{16\pi} \big( \sigma_{g}^{-1/2} \pm \sigma_{e}^{-1/2}\big)^2 / \big( 1 \pm g_c^{ge} \sigma_{g}^{-1/2}\sigma_{e}^{-1/2} \big) $, and $\sigma_{g(e)}=K_{g(e)}u_{g(e)}$ is the conductivity of the pure $g$ or $e$ orbital cloud. This exponent is shown in Fig.~\ref{fig:crit_exp} illustrating the typical behavior of the exponents as a function of the transverse confinement. When $\Delta_+^\kappa + \Delta_-^\kappa$ is larger than $\Delta_+^\sigma + \Delta_-^\sigma$ the orbital pairing state dominates the atomic density wave. This case is indicated by filled region in Fig.~\ref{fig:phase_diagYb}. and \ref{fig:phase_diagSr}. as a function of the transverse confinement $a_\perp$ and the population of the excited states. FFLO states have been studied extensively in various spin- and mass-imbalanced ultracold atomic systems, and their experimental realization is in progress \cite{Partridge06a, Zwierlein06a}, even in the one-dimensional case \cite{Liao10a}. 

In the third competing state the spin density oscillations show the slowest power low decay $|r|^{-\Delta_+^\kappa-\Delta_-^\kappa -(\Delta_+^{\tilde\sigma}+\Delta_-^{\tilde\sigma})/2 }$, where $\Delta_\pm^{\tilde\sigma}$ has the same form as $\Delta_\pm^\sigma$ above with the exchange of $\sigma_{g(e)}$ to $\tilde\sigma_{g(e)}=\tilde{K}_{g(e)}\tilde{u}_{g(e)}$, and the effective coupling $g_{g(e)}$ to $\tilde{g}_{g(e)}$. The corresponding Luttinger parameters are $\tilde{K}_{g(e)}= \sqrt{\frac{2\pi\hbar v_{g(e)} }{2\pi\hbar v_{g(e)} - 2 g_{g(e)} } }$, the sound velocities have the form $\tilde{u}_{g(e)}= \sqrt{(2\pi\hbar v_{g(e)} - g_{g(e)})^2 - g_{g(e)}^2}$, and the effective coupling is $2 \tilde{g}_c^{ge} =  g_{ge}^- - g_{ge}^+ $. The spin density wave is always suppressed by the atomic density wave. However, the spin density fluctuations can win over the orbital-FFLO close to the resonance at the electronic ground state scattering length. Therefore, in this regime (red region in the lower panel of Fig.~\ref{fig:phase_diagYb}. and \ref{fig:phase_diagSr}.) subdominant spin density wave order can be observed with dominant atomic density wave. 

The spin density wave order suppresses the orbital FFLO fluctuations typically close to the resonances, however, this dominance also occurs and becomes more emphasized for smaller Fermi velocities. Smaller Fermi velocity means smaller occupancy of the corresponding electronic state. The phase diagram depends on the population of the ground state and the excited state independently, not only the relative population of the two orbital states. For larger populations the resonances are narrower and well defined even for rather fine resolutions of the transverse oscillation length (e.g. in the order of $10^{-3} a_0$). As the Fermi velocities are decreasing the resonances start to show inner structure. Many narrow resonances occur and they are separated from each other with relatively larger distance as the population is decreased.  This situation relates to the case when the population of an orbital state becomes so small that the low energy physics of the system can not be described  with the particles in the linear environment of the Fermi points, instead the curvature of the spectra has to be taken into account. Since the linear approximation of the spectrum around the Fermi points is a key feature of our analysis, the case of the very small population is out of the reliability of the above results.

\begin{figure}
\includegraphics[scale=0.58]{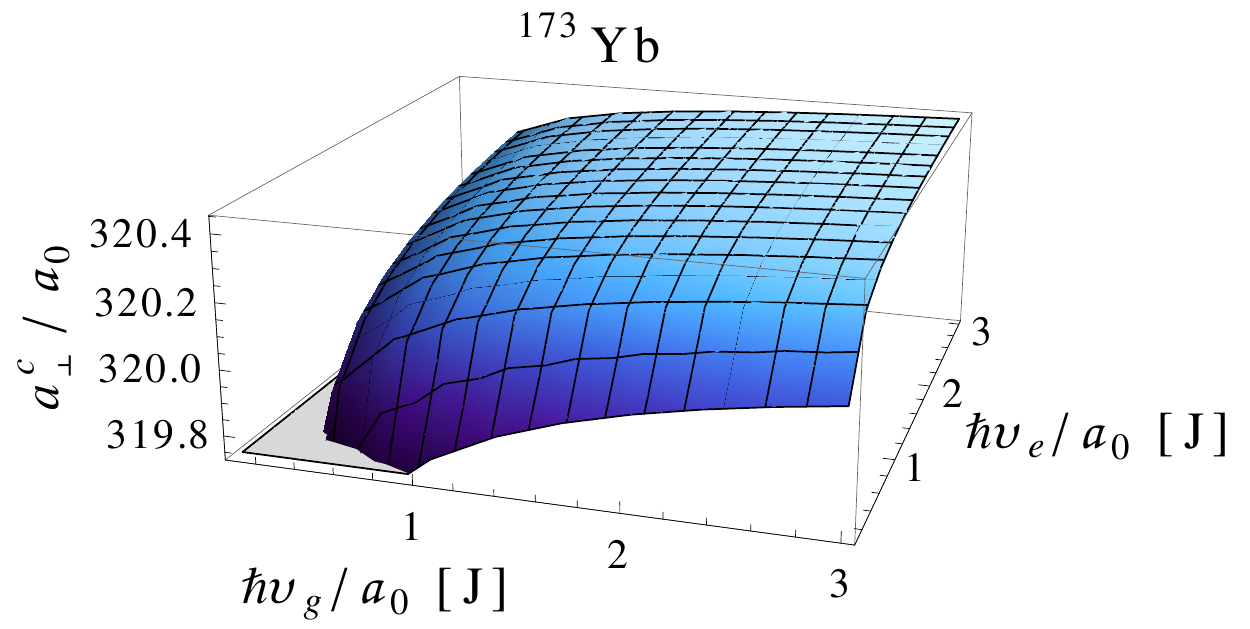}
\includegraphics[scale=0.58]{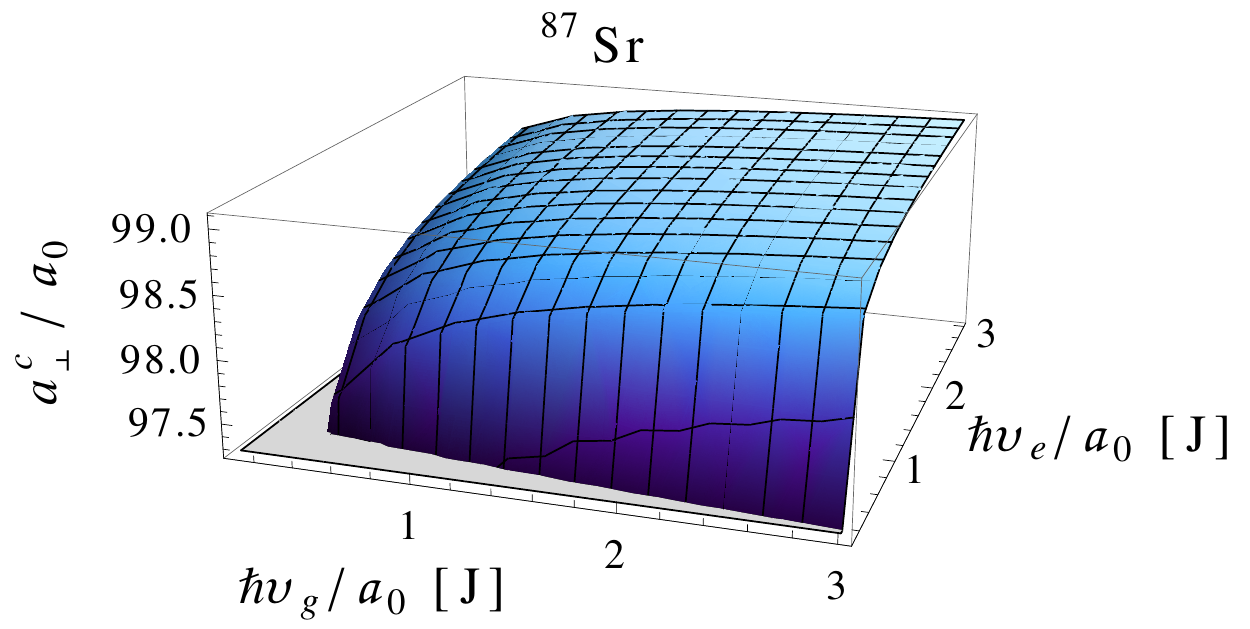}
\caption{(Color online) The critical transverse confinement as a function of the tunneling in the electronic ground state and excited state close to the resonance $a_\perp \approx 320.5\,\, a_0$ in case of $^{173}$Yb, and $a_\perp \approx 99.3\,\, a_0$ in case of $^{87}$Sr.}
\label{fig:amcrit}
\end{figure}

In a (effective) mass imbalanced system for sufficiently strong interactions the heavy and light particles prefer to occupy regions from which they exclude each other forming a domain structure and usually the heavy particles are compressed into a smaller region~\cite{Barbiero2010,Guglielmino2011,Roscilde2012,Sun2013,Partridge2006}.  Similar phenomena can be observed in our two-orbital system, too, when the compressibility  $\kappa_\pm=1/\sqrt{1\pm g_c^{ge} \sqrt{\kappa_g \kappa_e }}$ diverges, i.e. the denominator of $\kappa_\pm$ is zero. This can happen close to every resonances approaching them from both sides, and as the couplings reach sufficiently large value the competition of the various energy scales drives the system into a phase separated state. Fig.~\ref{fig:amcrit}. shows a typical behavior how the critical confinement $a_\perp^c$ is changing close to a resonance. In the figure we show the changing of $a_\perp^c$ as a function of the populations in the two orbital states around the specific resonance at $a_\perp\approx b\cdot 219.5 \,\, a_0 \approx 320.5\,\, a_0$ for $^{173}$Yb, and $a_\perp\approx b\cdot 68 \,\, a_0 \approx 99.3\,\, a_0$ for $^{87}$Sr. 

The initial confinement and populations of the orbital states determine the ground state of the system. One can prepare the cloud in a certain ground state far from the resonances, and that state can be driven into a segregated phase via a first order phase transition approaching a resonance. It means that the nature of the emerging segregated state can control via the preparation of the cloud. Phase separated states occur only for stronger couplings, therefore, their nature is difficult to catch. 
Nevertheless, based on our weak coupling analysis one can expect that the two slowest decaying states will form a domain structure in the segregated phases. Accordingly, three different states is expected to be realized in a possible experiment: (1) domains of orbital-FFLO and density wave alternate along the chain, (2) domains of orbital-FFLO and spin density wave alternate, and (3) domains of density wave and spin density wave alternate. 

In this paper we analyzed the possible ground states in a system of ultracold $^{87}$Sr or $^{173}$Yb atoms loaded into a one-dimensional chain. One part of the atoms are excited to a metastable excited state in order to mimic an additional internal (e.g. orbital) degree of freedom. Due to the electronic structure of the atoms the interaction does not depend on the spin, only the orbital state of the scattering atoms. Among others we showed using semiclassical approach that such a system is unstable against the emergence of exotic FFLO-like pairs consisting one atom in the electronic ground state and another one in the electronic excited state. These spin carrier pairs form a spin liquid-like state characterized by zero local expectation values of the pair operators with quasi-long-range correlations. We also showed that when the confinement length is tuned close to one of the characteristic scattering lengths, the system shows density instability and collapses into a phase separated state.

\section*{Acknowledgements}
I would like to thank L. Fallani, G. Pagano, and F. Scazza for the valuable discussions about some aspects of a possible experiment. I also thank  V. Gurarie, P. Lecheminant, G. Tak\'acs, and G. Zar\'and for the valuable discussions. This work was supported by the Hungarian National Research Fund (OTKA) No. K105149 and K100908, and the J\'anos Bolyai Scholarship of the Hungarian Academy of Sciences.

\bibliography{lowdim,ref}

\end{document}